\documentclass[aps,twocolumn,prb,amsmath,amssymb]{revtex4-1}
\usepackage[varg]{txfonts}
\usepackage{color}
\usepackage[colorlinks]{hyperref}

\usepackage{graphicx}
\usepackage{bm}

\def\XXint#1#2#3{{\setbox0=\hbox{$#1{#2#3}{\int}$}
    \vcenter{\hbox{$#2#3$}}\kern-.5\wd0}}

\def\be{\begin{equation}}
\def\ee{\end{equation}}

\def\bi{\begin{itemize}}
	\def\ei{\end{itemize}}
\def\bn{\begin{enumerate}}
	\def\en{\end{enumerate}}
\def\bea{\begin{eqnarray}}
\def\eea{\end{eqnarray}}
\newcommand{\bpm}{\begin{pmatrix}}
	\newcommand{\epm}{\end{pmatrix}}

\def\ba{\begin{array}}
	\def\ea{\end{array}}
\def\bd{\begin{displaymath}}
\def\ed{\end{displaymath}}

\renewcommand{\imath}{\hspace{1pt}\mathrm{i}\hspace{1pt}}

\renewcommand{\vec}{\mathbf}

\begin{document}

\title{Effects of dynamical noises on Majorana bound states}

\author{Roya Radgohar}
\affiliation{Department of Physics, Sharif University of Technology, Tehran 14588-89694, Iran}

\author{Mehdi Kargarian}
\email{kargarian@physics.sharif.edu}
\affiliation{Department of Physics, Sharif University of Technology, Tehran 14588-89694, Iran}

\begin{abstract}
The nonlocal nature of unpaired Majorana bound states (MBSs) in topological superconductors can be exploited to create topologically protected qubits and perform gate operations fault-tolerantly via braidings. However, the time-dependent noises induced by coupling to an environment which is inevitable in any realistic system could spoil the topological protection. In this work, we study the effects of various dynamical noises such as Lorentzian, thermal, and quantum point contact on the MBSs in the recently proposed one-dimensional topological superconductors. We begin by investigating the Kitaev $p$-wave superconductors and examine the effects of long-range hopping and pairing on the transition rate of MBSs. We found that, especially, the long-range pairings significantly reduce the transition rate of bound states. Then, we consider the recently discovered topological superconducting nanowires and magnetic chains. Our findings are consequential for the recent attempts to manipulate MBSs. In particular, for the latter two experimentally realized systems we argue how low magnetic/Zeeman fields and strong spin-orbit coupling make the MBSs more robust to noises.   
\end{abstract}

\maketitle

\section{Introduction}\label{Introduction}
Majorana bound states (MBSs) appear at the end of one-dimensional topological superconductors or in the vortex cores of two-dimensional chiral superconductors. Operationally a MBS is a fermionic quasiparticle that is its own antiparticle, i.e., $\gamma^{\dagger}=\gamma$. Therefore, the emergence of MBSs in a solid state system relies on equal superposition of electron and hole states, forming chargeless quasiparticles, and fermions with only one spin projection, e.g., the spinless fermions, are involved in the formation of Majorana states\cite{Alicea2011}. The one-dimensional spinless Kitaev superconductor with $p$-wave pairing potential is topologically nontrivial, supporting MBSs at the ends of open chain in the weak coupling regime\cite{Kitaev2001}. However, any material design of a one-dimensional superconductor requires lifting the spin degeneracy. 

\begin{figure}
  \centering
  \includegraphics[scale=0.3]{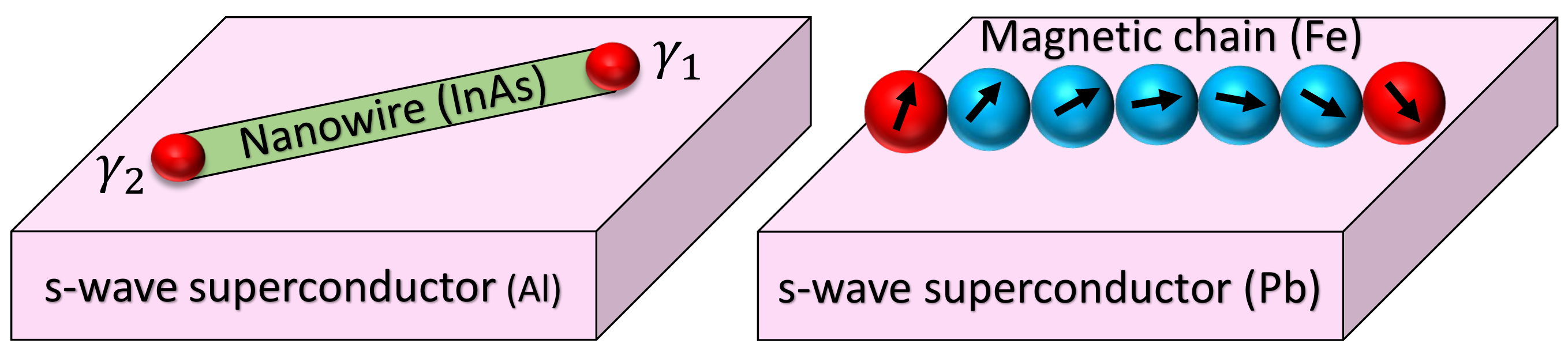}
  \caption{A schematic representation of experimental heterostructures of a semiconductor nanowire (left) and a magnetic chain (right) 
in proximity to an s-wave superconductor underneath. The Majorana zero modes $\gamma_{1}$ and $\gamma_2$ appear at the ends of the nanowire or magnetic chain when the induced superconductivity is in the topological phase.}
   \label{fig:system}
\end{figure}

Semiconductor heterostructures consisting of conventional materials such as nanowires with strong Rashba spin-orbit coupling in proximity to $s$-wave superconductors were proposed to exhibit nontrivial band topology, promising a hybrid structure supporting MBSs \cite{Oreg2010,Sau2010,Lutchyn2018}. The semiconductor heterostructure is shown schematically in the left panel of Fig.~\ref{fig:system}, where a nanowire of InSb (InAs) is grown on the surface of $s$-wave superconductor NbTiN (Al). Physically the strong Rashba coupling removes the spin degeneracy of electron states near the Fermi level, and a sizable Zeeman field can remove one of the energy bands. Hence, the single-particle states become effectively spinless giving rise to odd parity for pairing potential induced by the underneath superconductor. An observation of zero-bias conductance in hybrid structure signified the existence of MBSs at the ends of the nanowire \cite{Kouwenhoven:science2012}. Another hybrid structure consists of a ferromagnetic chain of iron atoms deposited on the surface of a superconductor\cite{NadjPerge2013,Klinovaja2013,NadjPerge2014,Li2014,Kim2018} as shown in the right panel of Fig.~\ref{fig:system}. The intrinsically ordered magnetic moments break the time-reversal symmetry, eliminating the need for an external magnetic field. Moreover, the angle between adjacent moments induces inter-spin component of hopping terms that mimics the effects of spin-orbit coupling. An observation of zero-bias tunneling conductance has been associated to MBSs \cite{NadjPerge2013}, though there are other explanations as well. 

Besides the fundamental importance of MBSs in our understanding of exotic quantum states, the surge of recent interests on MBSs originates in possible use of them to build topologically protected qubits and perform fault-tolerant quantum computation \cite{Nayak2008,Pachos2012}. The degenerate subspace of multiple MBSs provides a topological memory to store quantum information and a proper set of braidings of non-Abelian quasiparticles serves as gate operations on quantum states\cite{Ivanov2001}, all immune to local errors. 

Although, the topological qubits have some degree of robustness, especially against static disorder \cite{Ortiz2018}, they generically suffer from the time-dependent fluctuations of intrinsic properties of system as well as coupling to the environment. The latter coupling breaks the fermion parity-- an important ingredient of existence of MBSs-- through injection or removal of quasiparticles, giving rise to dynamic fluctuations that completely destroy coherence of Majorana qubits \cite{Goldstein2011}. Even the coupling to a parity-preserving reservoir such as finite-temperature bosonic bath can also destabilize MBSs, giving rise to an exponentially decay in correlation between MBSs \cite{Hu2015} and exposing braiding processes to errors \cite{Pedrocchi2015}. However, one can find a regime of parameters where there exists a long-lived quantum correlation between Majorana fermions in the presence of colored Markovian noise\cite{Hu2015}. For coupling to an Ohmic-like fermionic or bosonic bath with spectral density $\rho(\omega)\propto\omega^{Q}$, while the MBSs are robust in super-Ohmic regime $Q>1$, the coherence of zero modes is strongly suppressed in the Ohmic and sub-Ohmic regimes with $Q\leq1$ \cite{Ho:NJP2014}. The non-equilibrium noise effects coming from trijunction setups, despite conserving parity, decrease the coherence time of Majorana qubits \cite{Pedrocchi2015,Nag2019}. 

Since the noises are ubiquitous and indispensable in any physical system which could host MBSs, and hence, any successful protocol of quantum computation including initialization of qubits, implementation of gates, and readout is potentially subject to noises from various sources. Our paper is intended to investigate the effects of several noises on the robustness of MBSs and find the regime of parameters where the suffering effects of time-dependent noises are minimal. To this end, we focus on a class of noise sources relevant to the experimental setups inducing time-dependent fluctuations in chemical potential such as Lorentzian, thermal, and point contact noises. We identify the transition probability from zero-energy level to excited states as a measure for the fragility of MBSs against noises \cite{Konschelle2013} in three one-dimensional models: the $p$-wave Kitaev chain, Rashba nanowire, and magnetic chains, all in topological superconducting phases with MBSs at the ends of the chains. As discussed above, the last two models shown in Fig.~\ref{fig:system} are relevant to the current experimentally designed heterostructures, calling for determination of regimes of parameters where the effects of noises are minimal. For the Kitaev chain it is shown that the repulsive electron-electron interactions between nearest-neighbor sites decrease the decoherence rate \cite{Ng:scientificreports2015}, while the long-range many-body interactions between fermions reduce the lifetime of MBSs \cite{Wieckowski:PRL2018, Wieckowski:PRB2019}. 
We instead consider the effects of long-range tunnelings and superconducting pairings on transition probability. For the nanowire proximitized to the surface of an $s$-wave superconductor, the effects of strong Rashba coupling and Zeeman field on the robustness of bound states are studied. In particular we show that the stronger the former is, the more resilience against noises is achieved, a finding which could be important in looking for proper heterostructures with enhanced robust MBSs.   

The paper is organized as follows. In Sec.~\ref{sec:noise}, we introduce the noise models and the transition rate. We begin with a generalized version of the Kitaev chain with long-range hoppings and pairings in Sec.~\ref{sec:EKC} and numerically calculate the transition rate for MBSs. In Sec.~\ref{sec:SOC}, the effects of noises on MBSs in semiconductor nanowires in the presence of strong spin orbit coupling and magnetic field are presented, and in Sec.~\ref{sec:magnetic} the results for a magnetic atomic chain on the surface of superconductor are presented. We conclude in Sec.~\ref{sec:Conclusions}.

\section{Noise models}\label{sec:noise}
Before delving into the details of MBSs in one-dimensional systems and their resilience, in this section we introduce several noise models related to the hybrid structures and present a mathematical framework to calculate the transition probability of the MBSs to excited states. One of the sources of noise which is intrinsic to the electronic materials is the charge noise resulting from the quantum fluctuations of occupation numbers. This noise manifests itself as time-dependent fluctuations in chemical potential. The fluctuations in the electron spin states caused by the nuclear spin fluctuations is also another source of noise. 

Recent experiment shows that these two noise sources reveal a frequency spectrum \cite{Kuhlmann2013} that is described by a Lorentzian distribution function as  

\begin{equation}\label{Slorentz}
S_{\text{Lorentz}}(\omega)= S_0[1+(\omega-\omega_0)^2/(\delta\omega)^2]^{-1}, 
\end{equation}
where $\omega_0$ is the central frequency, $\delta\omega$ is the bandwidth, $S_{0}$ is the amplitude of the spectrum. In Sec.\ref{experiment} we give estimates for $S_0$ for all noises studied in this work. 
The limit of $\delta\omega\rightarrow 0$ recovers quasi-monochromatic frequency spectrum and the limit of $\delta\omega\rightarrow \infty$ corresponds to the quasi-white noise which contains equal contributions from all frequencies. We assume $\delta\omega=\Delta/\hbar$ with $\Delta$ being superconducting gap. This kind of noise spectrum has been used to describe the effects of an externally random fluctuating noise on physical systems\cite{Augello2010}. 

Besides the intrinsic noise sources described above, the thermal fluctuations are another source of noise. At non-zero temperature, the thermal fluctuations give rise to fluctuations in the occupation number of energy states and consequently in the chemical potential. In thermal equilibrium, the frequency spectrum of thermal noise is given by\cite{Schmidt2012,Konschelle2013} 

\begin{equation}\label{Sthermal}
S_{\text{Thermal}}(\omega)=S_{0}\text{exp}(-\hbar\omega/k_BT),
\end{equation}
where $k_B$ is the Boltzmann constant, $T$ is the temperature, and $\hbar$ is the reduced Planck constant.

The quantum transport across a quantum point contact (QPC) between a superconductor and a semiconductor or magnetic atomic chain in hybrid structures suffers from a non-equilibrium electrical current noise known as shot noises. The latter is a consequence of random transfer of quantized charged carriers through mesoscopic conductors. If the energy of an electron impinging on the surface of superconductor is smaller than the superconducting gap ($E<\Delta$), it can be Andreev reflected, through which a hole is reflected back to the semiconductor and a Cooper pair with charge $2e$ is injected to superconductor. The reverse process is also possible where a Cooper pair recombines with a hole in the semiconductor and produces an electron. In equilibrium, both processes occur with equal probability, leading to no net current flow. Hence, a bias voltage ($V$) across the junction of semiconductor-superconductor is required to achieve a finite current flow. Using the scattering theory, the frequency spectra of the shot noises for both cases have been computed in Ref.[\onlinecite{Blanter2000}]. At zero temperature, they are as follows:

\begin{eqnarray}
&&S_{eq}(\omega)=\frac{2e^2\omega}{\pi}\sum_n D_n,\\
&&S(\omega)=\frac{2e^2\omega}{\pi}\sum_nD_n^2+\frac{4e^3V}{\pi\hbar}\sum_{n}D_n(1-D_n),
\end{eqnarray}
where $S_{eq}$ is the shot noise in equilibrium ($V=0$), and $S$ is the shot noise at finite voltage ($\hbar\omega<eV$). Here  $D_n=T_n^2(2-T_n)^{-2}$, where $T_n$ is the $n$th transmission eigenvalue between the interface and semiconductor. The difference $S(\omega)-S_{eq}$ is used to characterize the QPC noise (also called excess noise) as\cite{Blanter2000}

\begin{equation}\label{Sqpc}
S_{\text{QPC}}(\omega)=S_0\left(1-\frac{\hbar\omega}{eV}\right),
\end{equation}
where $S_0=(2e^3V/\pi\hbar)\sum_{n}D_n(1-D_n)$ .
\par

Having introduced several dynamical noise spectra in equations \eqref{Slorentz}, \eqref{Sthermal}, and \eqref{Sqpc}, we now discuss how the effects of latter  noises on the MBSs are taken into account, which is the main subject of this work. We also ignore other sources which could lead to fluctuations in spin-orbit interactions \cite{Kuhlmann2013} and superconducting pairings \cite{DeVisser2014,Brihuega2011,Petkovic2013}. Following Ref.[\onlinecite{Konschelle2013}], we assume that the dynamical noises perturb the chemical potential 
as $\mu(t)=\mu+\zeta f(t)$, where $\mu$ is the unperturbed chemical potential, $\zeta$ is the coupling constant, and $f(t)$ is the interacting potential amplitude encoding the information about the type of noise under consideration. The latter term perturbs the Hamiltonian as $H=H_0+\zeta f(t)\bf{M}$, where the unperturbed Hamiltonian is given by $H_0$, and $\bf{M}$ is a density operator associated with the change in the chemical potential which will be specified for our models in next sections. Let us denote the zero-energy state by $|0\rangle$ and the excited states by $|q\rangle$. The transition probability out of $|0\rangle$ is given by
\begin{equation}
P(t)\equiv\sum_{q} |\langle q|U(t)|0\rangle|^2,
\end{equation}
where $U(t)$ is the time-evolution operator which will be specified shortly. Assuming the coupling $\zeta$ is small, we may apply the first order time-dependent perturbation theory to obtain the following expression for the time-evolution operator:
\begin{equation}\label{eq:Evolution}
U(t)\approx U_0(t)+\frac{\zeta}{i\hbar}\int_{0}^{t}U_{0}^{\dag}(\tau)
f(\tau)\textbf{M}U_0(\tau)d\tau,
\end{equation} 
where $U_0(t)=e^{-itH_0/\hbar}$. Since we are interested in averaged time evolution of the system, we obtain the average probability $\bar{P}$ 

\begin{eqnarray}
\bar{P}(t)&=&\frac{\zeta^2}{\hbar^2}\sum_q\int_{0}^{t}\int_{0}^{t}d\tau d\tau^{\prime}\langle f(\tau)f(\tau^{\prime})\rangle\langle q|U_0^{\dag}(\tau){\bf{M}}U_0(\tau)|0\rangle\cr\cr
&\times&\langle 0|U_0(\tau^{\prime}) {\bf{M}}^{\dag}U^{\dag}_0(\tau^{\prime})|q\rangle
\end{eqnarray}
where the noise correlation function $\langle f(\tau)f(\tau^{\prime})\rangle$ is related to the frequency spectrum of noise $S(\omega)$ in equations \eqref{Slorentz}, \eqref{Sthermal}, and \eqref{Sqpc}, as \cite{Schoelkopf2003}

\begin{equation}
\langle f(\tau)f(\tau^{\prime})\rangle =\int\frac{d\omega}{2\pi} e^{i\omega(\tau^{\prime}-\tau)}S(\omega).
\end{equation}

Finally, the probability rate is given by the time-derivative of $\bar{P}$. It reads as 
\begin{eqnarray}
\Gamma\equiv\frac{d\bar{P}}{dt}=\frac{\zeta^2}{\hbar^2}\sum_{q} |\langle q|\textbf{M}|0\rangle|^2 
\int d\omega S(\omega)\delta\left(\omega-\varepsilon_{q}/\hbar\right),
\label{dPdt}
\end{eqnarray}
where $\varepsilon_{q}$ is the eigenenergy of $q$th-excited state. For simplicity, we take $ \zeta^2 S_0/\hbar^2=1$ and define $\omega_D=\Delta/\hbar$ ($=\delta\omega$) as a frequency associated to superconducting gap. We note that this equation is similar to the Fermi's golden rule for transition rates \cite{Schoelkopf2003}. In the following sections we use Eq.\eqref{dPdt} to evaluate the effect of various noise sources on the MBSs.  

\section{Kitaev $p$-wave chain with long-range hoppings and pairings \label{sec:EKC}}
The simple theoretical model satisfying both conditions of equal superposition of electron and hole states and having only one spin species  is the Kitaev chain introduced in Ref. [\onlinecite{Kitaev2001}]. The model is composed of spinless fermions with nearest-neighbor tunnelings and superconducting pairings. In the weak coupling regime the bulk states are topological and MBSs appear at the ends of an open chain. While the original model has short-range hopping and pairing amplitudes, the recent theoretical and experimental works have generalized the Kitaev chain to include long-range interactions\cite{Niu2012,Vodola2014,Dutta2017,Alecce2017}. Our main objection in this section is to study the influence of long-range interactions in Kitaev chain on the sensitivity of MBSs when subjected to noise sources introduced in preceding section. 

\begin{figure*}[!htb]
  \centering
  \includegraphics[scale=0.5]{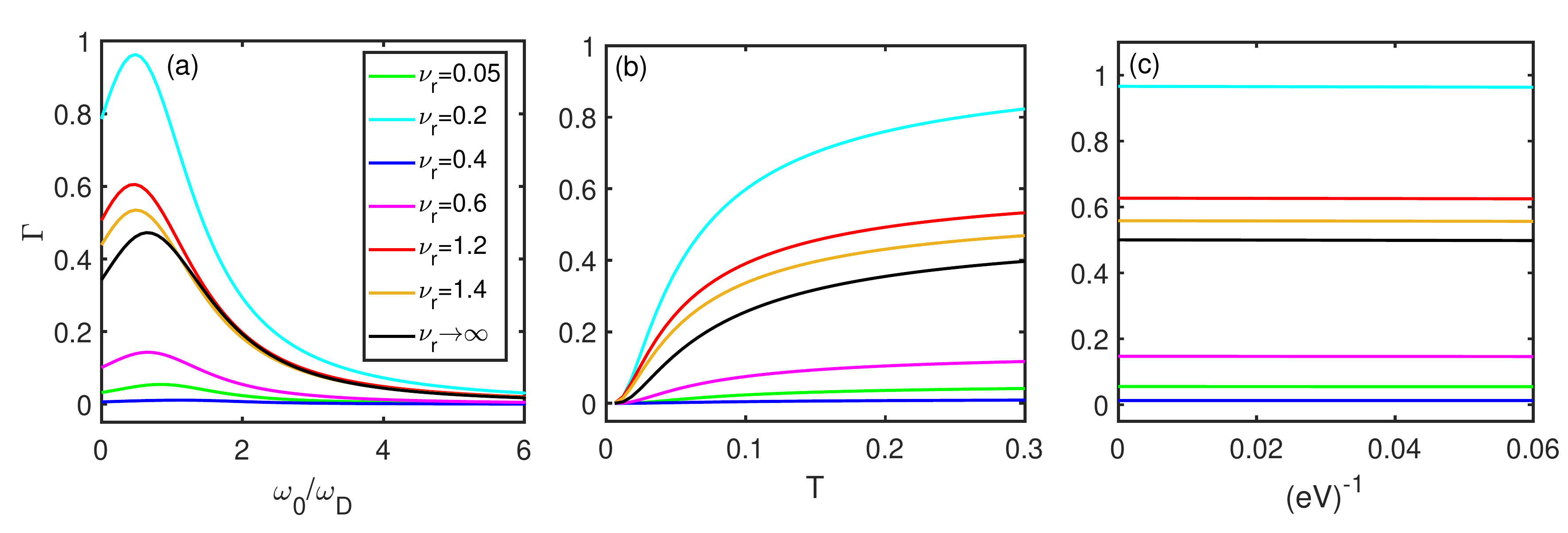}
  \caption{The probability rate to excite MBSs in an extended Kitaev chain with long-range hopping in the presence of (a) Lorentzian, (b) thermal, and (c) QPC noises. The parameters used are $N=101$, $J_0=1$, $\mu=-1$ and $\Delta_0=0.1$. The value of $\nu_{r}$ determines the strength of long-range hoppings; $\nu_{r}\rightarrow\infty$ corresponds to the original Kitaev model.}
    \label{fig:ExtendedHopping}
\end{figure*}

The generalized Kitaev chain is obtained by letting hopping and pairing amplitudes to extend to $r$-th and $s$-th neighbors, respectively. The Hamiltonian reads as

\begin{eqnarray}
\label{EKCFH}
H_{0}&=&-\sum_{j=1}^{N}\mu\left(a^{\dag}_ja_j-\frac{1}{2}\right)
-\sum_{l=1}^{r}\sum_{j=1}^{N-l}\left(J_la^{\dag}_ja_{j+l}+h.c.\right)\cr
&+&\sum_{l=1}^s\sum_{j=1}^{N-l}\left(\Delta_la_ja_{j+l}+h.c.\right),
\end{eqnarray}
where $\mu$ is chemical potential, $N$ denotes total number of sites, and $a_{j}(a_{j}^{\dag})$ is a fermionic annihilation (creation) operator. Moreover, the strength of long-range hoppings and pairings decreases with the distance between sites as power law functions $J_l=J_0 l^{-\nu_r}$ and $\Delta_l=\Delta_0 l^{-\nu_s}$, respectively, where $J_0$ and $\Delta_0$ are the corresponding nearest-neighbor values and the exponents of $\nu_s$ and $\nu_r$ control the strength of amplitudes so that $\nu_s,\nu_r<1 (\nu_s,\nu_r>1)$ correspond to long (short)-range interactions. Taking the limit $\nu_r,\nu_s \rightarrow\infty$, the original Kitaev model is recovered. 

The model \eqref{EKCFH} can be simulated in cold atomic gases interacting through tunable Feshbach resonance \cite{Gurarie2007,Jiang2011,Buehler2014} or in a setup of planar Josephson junctions in proximity to a 2D electron gas where long-range pairings and hoppings are controlled experimentally \cite{Liu2018}. The phase diagram of the Hamiltonian \eqref{EKCFH} contains topological superconducting phases with MBSs \cite{Alecce2017}. 

The model is much easier to analyze in Majorana representation of fermion operators. The transformation reads as

\begin{equation} 
a_j=\frac{1}{2}(c_{2j-1}+ic_{2j})  \quad a^{\dag}_{j}=\frac{1}{2}(c_{2j-1}-ic_{2j}),
\label{FMP}
\end{equation}
where the Majorana operators satisfy the Clifford algebra $\{c_i,c_j\}=2\delta_{i,j}$ for $i,j=1,\ldots,N$. The Hamiltonian becomes 

\begin{eqnarray}\label{eq:HMKitaev}
H_{0}=&-&\frac{i\mu}{2}\sum_{j=1}^{N}c_{2j-1}c_{2j}
+\frac{iJ_{0}}{2}\sum_{l=1}^{r}\sum_{j=1}^{N-l}\frac{1}{l^{\nu_r}}\left( c_{2j}c_{2(j+l)-1}-c_{2j-1}c_{2(j+l)}\right)\cr
&+&\frac{i\Delta_0}{2} \sum_{l=1}^{s}\sum_{j=1}^{N-l}\frac{1}{l^{\nu_s}} \left(c_{2j-1}c_{2(j+l)}+c_{2j}c_{2(j+l)-1}\right).
\end{eqnarray}

We consider a finite open chain with odd number of sites $N$ and $r=s=(N-1)/2$, and work in a regime of parameters where the model is in a topological superconducting phase. For the sake of simplicity and arguments we consider two cases separately: 1) the case of nearest-neighbor pairing and long-range hoppings is studied in Sec.~\ref{subsec:Hopping}, and 2) the case of nearest-neighbor hopping and long-range pairings is discussed in Sec.~\ref{subsec:Pairing}. To connect to our discussions of noises in the preceding section Sec.~\ref{sec:noise}, a dynamical shift in chemical potential $\mu\rightarrow \mu+\zeta f(t)$ in Eq.~\eqref{eq:HMKitaev} yields $\textbf{M}=\sigma^y$, where $\sigma^y$ is the Pauli matrix in Majorana basis for spinless fermions. Note that for models with spins, as discussed in next sections, $\textbf{M}=\sigma^x$; see Apendices for details. Using Eq.~\eqref{dPdt} and eigenvectors and eigenvalues of the Hamiltonian, we evaluate the transition rate $\Gamma$ of MBSs.

\subsection{Kitaev chain with long-range hopping \label{subsec:Hopping}} 
For this case the hopping terms are long-ranged, i.e., $\nu_{r}$ in Eq.~\eqref{eq:HMKitaev} is finite, but the nearest-neighbor pairing is obtained by taking the limit $\nu_s\rightarrow\infty$. This model has a rich phase diagram studied in Ref.[\onlinecite{Alecce2017}]. The topological phase is characterized by the nontrivial winding numbers $w=\pm1$ in the following regime of parameters:
\begin{equation}
-2J_0\sum_{l=1}^{N-1}\frac{1}{l^{\nu_r}}<\mu<2J_0\sum_{l=1}^{N-1}\frac{(-1)^{l+1}}{l^{\nu_r}}.
\end{equation}

In this regime the chain hosts Majorana modes localized at the ends. The transition probabilities of MBSs affected by distinct types of noises are shown in Fig.~\ref{fig:ExtendedHopping}. In all panels the black solid curve corresponds to the behavior of $\Gamma$ in the original Kitaev model obtained by $\nu_r\rightarrow\infty$. Therefore, the plots provide insights on how the range of hopping affects the transition. 

The first panel exhibits the behavior of $\Gamma$ versus the central frequency $\omega_{0}/\omega_{D}$ of the Lorentzian noise in Eq.~\eqref{Slorentz}. For all range of hoppings $\nu_r$, a peak appears for $\omega_0/\omega_D<1$, which is attributed to the resonance with superconducting gap. It is seen that the strength of the long-range hopping can significantly affect the transition probability. In the regime of short-range hopping interactions ($\nu_r>1$), the probability rate surpasses the corresponding values of the original Kitaev model (black curve), and by further increase of $\nu_{r}$ the curves approach the latter model. In the long-range hopping regime, where  $\nu_r<1$, the probability rate shows a totally different behavior. For values around $\nu_{r}=0.2$, the $\Gamma$ is quite large, while for $\nu_{r}=0.4$ is exceedingly small. An inspection of Eq.~\eqref{dPdt} shows that two factors conspire to determine the probability rate: the transition matrix element $\langle q|\textbf{M}|0\rangle|^2$ and the accumulation of states whose energies $\varepsilon_{q}$ are close to $\omega_0$. The latter makes $S_{\text{Lorentz}}$ quite appreciable for many states. 

We found that for example for $\nu_{r}=0.2$ the matrix element is rather large for states near the energy gap. Also, the gap in the energy spectrum is small, and therefore many states $|q\rangle$ contribute to the noise spectrum which is detrimental in having small values of $\Gamma$.
For other values of long-rang hopping, say $\nu_{r}=0.4, 0.6$, the gap in the spectrum pushes many states away from MBSs, suppressing the transition probability rate. On the other hand, for $\nu_r>1$ the superconducting energy gap is relatively large, so less states are involved in the noise spectrum, and since the $|\langle q|\textbf{M}|0\rangle|^2$s become rather large for states near the gap, a relatively large value of $\Gamma$ arises. And, in the limit of $\nu_r\rightarrow\infty$ the original Kitaev model is reached out.  

\begin{figure*}[!htb]
  \centering
  \includegraphics[scale=0.5]{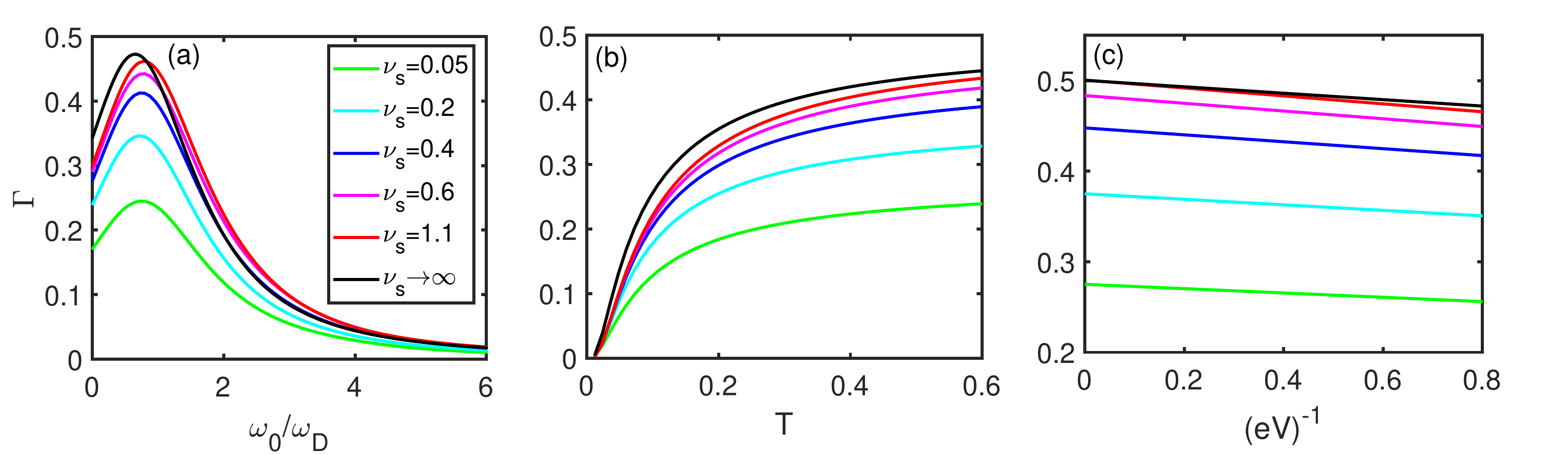}
  \caption{The probability rate to destroy MBSs in a Kitaev chain of length of $N=101$ with long-range pairing terms affected by (a) Lorentzian, (b) thermal, and (c) QPC noises. The parameters are $J_0=1$, $\mu=-2$, $\Delta_0=0.1$. The value of $\nu_{s}$ determines the strength of long-range pairings; $\nu_{s}\rightarrow\infty$ corresponds to the original Kitaev model.}
    \label{fig:Extendedpairing}
\end{figure*}

The results for thermal and QPC noises are shown in Fig.~\ref{fig:ExtendedHopping}(b) and (c), respectively. The transition rate increases at high temperatures, since the thermal weight in Eq.~\eqref{Sthermal} is relatively large for many states and they contribute in $\Gamma$. However, it turns out at least for some ranges of small values of long-range hopping strength $\nu_r$ the transition rate can be significantly suppressed. For the QPC noise the values of transition rate do not change with gate voltage, however, again it's seen that there exists a window of $\nu_r$ where the transition rate is decreased substantially. We note that the general behavior of probability rate with $\nu_r$ is similar for all three noise sources.

\subsection{Kitaev chain with long-range pairing\label{subsec:Pairing}}
Next we move to the second case of non-local superconducting pairing amplitudes given by finite value of $\nu_{s}$ in Eq.~\eqref{eq:HMKitaev}, while the hopping amplitudes are restricted to nearest-neighbor sites. Again note that the limit $\nu_s\rightarrow\infty$ recovers the original Kitaev model. The model with finite $\nu_{s}$ exhibits a nontrivial topological phase in a parameter range of $-2<\mu/J_0<2$\cite{Alecce2017}. 

The results of transition rate for three noise sources are shown in Fig.~\ref{fig:Extendedpairing}. Again the black curves in all panels show the variation of $\Gamma$ for the original Kitaev model. It is clearly seen that for all types of sources the transition rate for the latter model lies at the upper limit of curves. A striking feature of these plots is that as the strength of the long-range pairing is increased by decreasing $\nu_{s}$, the transition probability rate is reduced. The reason can be traced back to the energy gap, the number of energy states close to the zero-energy state, and the matrix elements as discussed above. Indeed, for this case the energy gap is increasing smoothly with increasing $\nu_s$, while the values of matrix elements remains rather small. Both effects then cooperate in yielding comparatively small values for $\Gamma$. Having established such a unique behavior, the results show that the harmful effects of noises on MBSs can be reduced in systems with long-range superconducting pairing amplitudes.

\section{nanowires in proximity to an $s$-wave superconductor \label{sec:SOC}}
In this section we present the results for noise on the MBSs in one of the most realistic and experimentally realized platforms. A schematic of the model is shown in the left panel of Fig.~\ref{fig:system}(a). The system is a heterostructure of a semiconductor nanowire in proximity to an  
$s$-wave superconductor. The role of latter superconducting substrate is to induce pairing potential into the nanowire. The main microscopic ingredients to have a topologically nontrivial pairing gap in the nanowire are strong spin-orbit coupling and a moderate Zeeman field, which is provided by a magnetic field \cite{Oreg2010,Sau2010, Lutchyn2018}. The heterostructure has been designed experimentally with strong evidence of the existence of MBSs appearing at the open ends of the nanowire \cite{Kouwenhoven:science2012}. Our objection is to investigate the effects of noises on the MBSs and determine the range of parameters where the latter states remain less influenced by noises.    

The continuum model Hamiltonian capturing the main physics of topological superconductor in this heterostructure is \cite{Lutchyn2010,Lutchyn2018}

\begin{eqnarray}
H&=&\sum\limits_{\lambda,\lambda^{\prime}}\int_{0}^{L}dx 
\psi_{\lambda}^{\dag}(x)\left(-\frac{\hbar^2\partial_{x}^2}{2m^*}-
\mu+i\alpha\hat{\sigma}_y\partial_x+h\hat\sigma_x\right
)_{\lambda\lambda^{\prime}} \psi_{\lambda^{\prime}}(x)\cr
&&+\Delta \int_0^L dx \left(\psi^{\dag}_{\uparrow}(x)
\psi^{\dag}_{\downarrow}(x)+h.c.\right),
\end{eqnarray}
where $m^*$ and $\mu$ are the effective mass and chemical potential, respectively. The third term describes the Rashba spin orbit coupling (RSOC) in semiconductor nanowire which lifts the spin degeneracy. The Zeeman energy $h=g\mu_B B$, where $g$ is the Lande g-factor and $\mu_B$ is the Bohr magneton, opens a gap in the energy spectrum. A strong enough magnetic field can push one of the bands above the Fermi level, and therefore creates single-degenerate electron state near the Fermi level.

\begin{figure*}[!htb]
  \centering
  \includegraphics[scale=0.5]{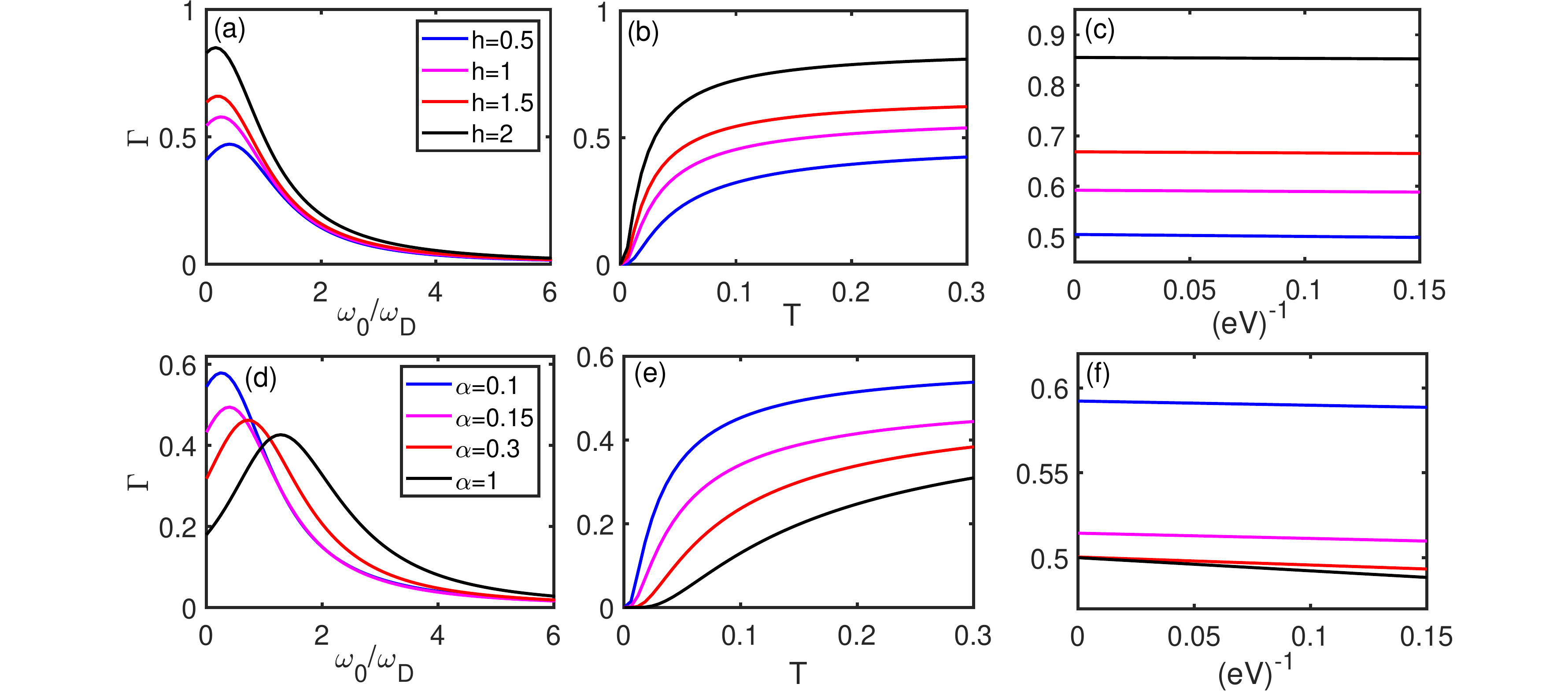}
  \caption{The probability rate of MBSs in a Rashba nanowire with $N=401$. Different solid colored curves in top (bottom) row are for different values of magnetic field $h$ (spin-orbit coupling $\alpha$). The panels are for (a,d) Lorentzian, (b,e) thermal, and (c,f) QPC noises. The parameters are $J=1$, $\mu=-2$ and $\Delta=0.1$. In top panels $\alpha=0.1$ and in bottom panels $h=1$.}
    \label{fig:KitaevSOChAlpha}
\end{figure*}

Since we are interested in full spectrum of an open chain, in the following we use the corresponding Hamiltonian on a lattice. The Hamiltonian reads as \cite{Mazziotti2018}

\begin{eqnarray}
H&=&-\mu\sum_{\lambda,j=1}^{N}a^{\dag}_{j,\lambda}
a_{j,\lambda}-J\sum_{\lambda,j=1}^{N-1}\left(a^{\dag}_{j,\lambda}
a_{j+1,\lambda}+h.c.\right)\cr
&&+\alpha\sum_{\lambda,\lambda^{\prime},j=1}^{N-1}\left
[i\sigma^y_{\lambda,\lambda^{\prime}}\left(a^{\dag}_{j,\lambda}
a_{j+1,\lambda^{\prime}}-a_{j+1,\lambda}^{\dag}a_{j,\lambda^{\prime}}\right)\right]\cr
&&-h\sum_{\lambda,\lambda^{\prime},j=1}^{N} a^{\dag}_{j,\lambda}\sigma_{\lambda,\lambda^{\prime}}^{x}a_{j,\lambda^{\prime}}
+\Delta\sum_{j=1}^{N}\left(a^{\dag}_{j,\uparrow}a^{\dag}_{j,\downarrow}+h.c.\right).
\label{SOCMajorana}
\end{eqnarray}

Representation of this Hamiltonian in terms of the Majorana fermions is given in Appendix \ref{app:SOC}. When $h>\sqrt{\Delta^2+(\mu+J)^2}$, the induced superconducting state is topologically nontrivial \cite{Lutchyn2018,Mazziotti2018} and the MBSs appear. In the topological phase we numerically diagonalize the Hamiltonian on a finite open system and use Eq.\eqref{dPdt} to compute the transition rate. In particular, we would like to find a regime of  parameters $h$ and $\alpha$ where the MBSs are relatively immune to dynamical noises. The spin-orbit coupling $\alpha$ can be changed by utilizing material combinations having different Lande $g$ factor and effective electron mass \cite{Lutchyn2018}. For example for the heterostructure InAs/Al  $\alpha_{\text{expr}}=0.2 - 0.8~ \text{eV}.\AA$, while for InSb/NbTiN it is $\alpha_{\text{expr}}=0.2 - 1~\text{eV}.\AA$.

In the following we work in a regime of parameters $h$ and $\alpha$ where the topological superconducting phase sets in and there exist MBSs at the ends of open chain. The results for transition probability are shown in Fig.\ref{fig:KitaevSOChAlpha}. In the first row of panels we examine the effects of the magnetic field on the transition rate $\Gamma$ with $\alpha=0.1$. For the Lorentz noise shown in Fig.\ref{fig:KitaevSOChAlpha}(a), while for all values of magnetic fields MBSs appear as zero-energy modes, it's desirable to work in the regime of weak magnetic field so that the MBSs are less impacted by dynamic fluctuations. We found that by increase of magnetic field the gap in the spectrum becomes smaller and, consequently, many states are involved in the transition probability. In fact, the number of energy states with non-zero transition matrix elements decreases with field, but the value of transition matrix elements is increased significantly. Therefore, the large field limit exposes the MBSs to noise and rises the transition probability. The thermal noises become more prominent at high temperatures and, as shown in Fig.\ref{fig:KitaevSOChAlpha}(b), small values of magnetic field can suppress the transition rate. Similar effects of suppression of transition rate by decreasing magnetic field are also demonstrated for the QPC noise in Fig.\ref{fig:KitaevSOChAlpha}(c).   

Another very important parameter as discussed above is the spin-orbit coupling $\alpha$. The results are shown in the second row of panels in Fig.\ref{fig:KitaevSOChAlpha}. For the Lorentzian noise, the transition rates for several values of $\alpha$ are shown in Fig.\ref{fig:KitaevSOChAlpha}(d). The results indicate that for small values of $\alpha$ the transition rate becomes large for lower part of spectrum. Our detailed analysis show that for small $\alpha$, despite having a small gap, the matrix element is large for excited states near the gap giving rise to a large transition rate. It starts diminishing by increasing $\alpha$ within the low-frequency window of noise spectrum. For larger frequencies, however, the rise of matrix elements leads to increment of transition rate. The results for thermal and QPC noises are shown in Fig.\ref{fig:KitaevSOChAlpha}(e) and (f), respectively. Now we see that the noise effects are substantially diminished by increasing $\alpha$, and thus the transition rate is decreased. These results show that choosing nanowires with large spin-orbit interaction will make the MBSs more immune to noises. 

\begin{figure*}[!htb]
	\centering
	\includegraphics[scale=0.5]{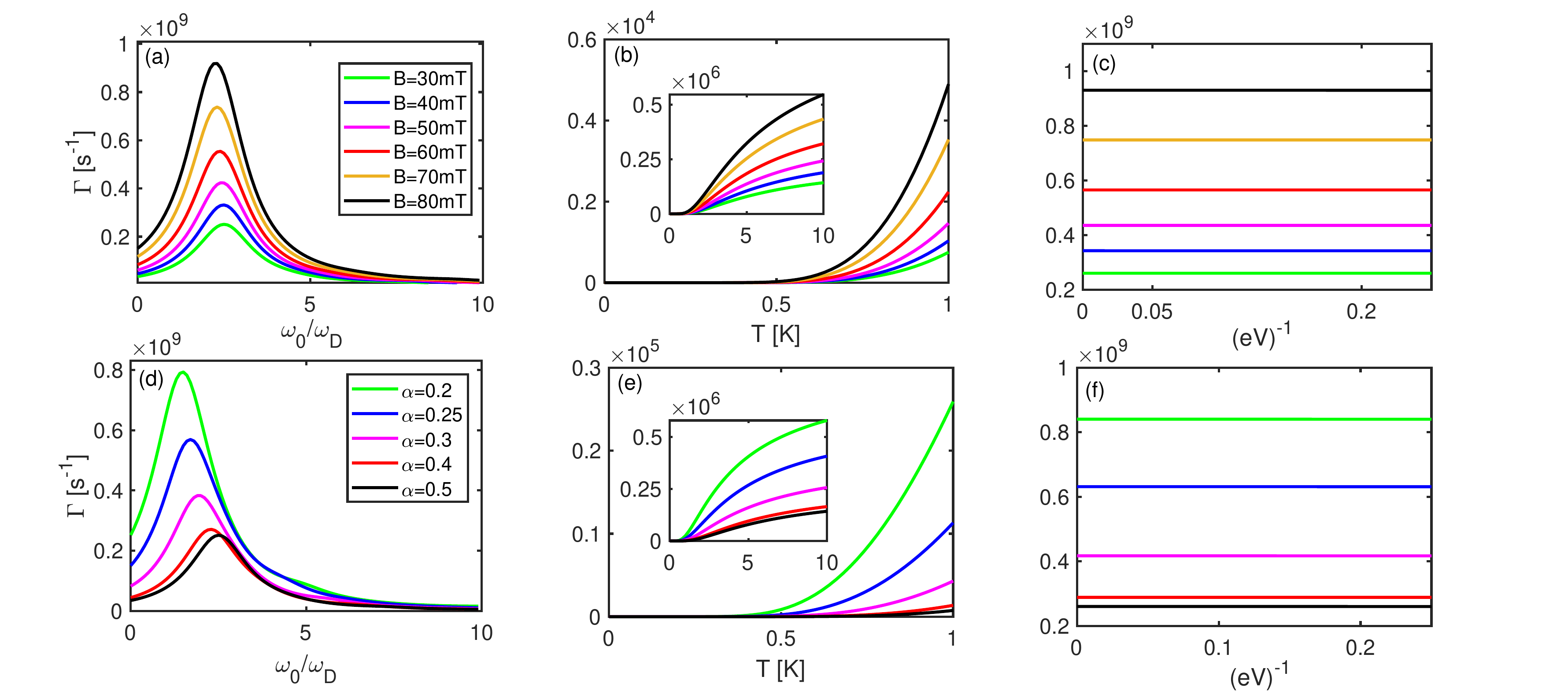}
	\caption{The transition rate of MBSs in nanowire InAs/Al heterostructure. Different solid colored curves in top (bottom) row are for different values of magnetic field $B$ (spin-orbit coupling $\alpha$). The panels are for (a,d) Lorentzian, (b,e) thermal, and (c,f) QPC noises. The parameters are $g=8$, $\Delta=0.2$~meV. In top panels $\alpha=0.5~\text{eV}.\AA$ and in bottom panels $B=30~\text{mT}$. The insets in middle panels show the rate for a large range of temperatures.} 
	\label{fig:estimate}
\end{figure*}

\subsection{Evaluation of transition rates for InAs/Al heterostructure \label{experiment}}                  
Now we are in a good position to evaluate the transition rates for MBSs realized in InAs/Al nanowire heterostructure shown schematically in Fig.~\ref{fig:system}. In order to compute $\Gamma$ we have to evaluate amplitude of noise spectrum $S_0$ and also the prefactore $\zeta^2S_0/\hbar^2$ in Eq.\eqref{dPdt}. The Lorentzian noise source exhibits random telegraph signal (RTS) with the following noise power spectral density \cite{Longoni1995} 
\begin{eqnarray}\label{Lorentztrap}
S(\omega)=2(\Delta I)^2\frac{n n_t}{(n+n_t)^2}\frac{1/v_{th}s_n(n+n_t)}{1+\omega^2/(v_{th}s_n(n+n_t))^2},
\end{eqnarray}
where $\Delta I$ is the amplitude of the RTS waveform, $v_{th}$ is the thermal velocity of carriers, and $s_n$ is the trap's electron capture cross-section. Here, $n$ is the mobile electron density around the trap location, and $n_t=N_c\exp(-\frac{E_c-E_t}{k_{B}T})$ is the effective electron density associated with the energy level of trap $E_t$. Also, $N_c$ and $E_c$ are the electron density and energy of conduction band. Upon the comparison with Eq.\eqref{dPdt}, the value of $S_0$ can be estimated as $S_0=2(\Delta I)^2\frac{1}{v_{th}s_n}\frac{nn_t}{(n+n_t)^3}$.

To evaluate $\zeta^2S_0/\hbar^2$ we note that for the Lorentzian noise, $\zeta=e/(\Delta G)$, where $\Delta G=e\mu_e N_D A/L$ is conductance change of nanowire and $N_D$ is density of donors. Also, $\mu_e$ denotes the mobility of electrons, and $A$ and $L$ are cross section and length of nanowire, respectively, and $\Delta I=e\mu_eV_a/L^2$, where $V_a$ is the applied voltage. The parameters adapted for InAs nanowire are: $L=100~\text{nm}$, temperature $T=20~\text{mK}$, activation energy $E_c-E_t\simeq 0.1~\text{meV}$, $V_a\simeq 1\text{meV}$, $n\simeq10^{15}~\text{cm}^{-3}$ and $N_D\simeq 10^{14}~\text{cm}^{-3}$ \cite{Lutchyn2018,Alzamil2011}. Therefore, we obtain $\zeta^2S_0/\hbar^2=\frac{2e^2V^2}{\hbar^2N_D^2A^2L^2}\frac{nn_t}{v_{th}s_n(n+n_t)^3}\simeq 10^{9}\text{s}^{-1}$. 

 For the thermal noise, $S_0=2m^*_g\omega l^2/8\epsilon_F$ and $\zeta=e^2/4\pi\epsilon\epsilon_0 d$, where $m^*_g\simeq 0.1m_e$ is the effective mass of electrons, $\epsilon_F\simeq 1~\text{eV}$ is the Fermi energy in the gate, and $\epsilon_0$ is the vacuum permittivity. Other parameters are dielectric constant $\epsilon\simeq 10$, width of Gaussian profile $l\simeq d $ and the typical gate-wire distance $d$ \cite{Konschelle2013}. It yields $\zeta^2S_0/\hbar^2\simeq 10^{6} \text{s}^{-1}$.

Finally for the case of QPC noise, $\zeta=e/G$ with $G=e^2 N_{\perp}l_{m}/2\pi\hbar L$, where $ N_{\perp}$ denotes the number of transverse conducting channels and $l_{m}$ is the mean free path of electrons. We use typical values as  $N_{\perp}\sim 300$, $V=1\text{V}$, $l_m=200~\text{nm}$, $L=100~\text{nm}$ and $T_n=0.6$. We then obtain $\zeta^2S_0/\hbar^2\simeq 10^{9}\text{s}^{-1}$.
  
In Fig.~\ref{fig:estimate} we summarize the estimate of transition rate $\Gamma$ for three dynamical noises. The parameters in all plots are within the experimental reach. The results are highly suggestive that MBSs remain less affected by noises at low temperatures, weak magnetic fields, and strong spin-orbit coupling. Also one can see that the Lorenztian, panels (a) and (d), and QPC noises, panels (c) and (f), produce a large degree of transition rate, much larger than the thermal noise in panels (b) and (e).  

Therefore, any protocol for creating and manipulating of MBSs has to be designed in a way to mitigate the effects of such noises significantly. In particular, our results for dynamical noises imply that the latter may desctructively affect gate operations in topological quantum computation protocols due to decoherence and overwhelmed excitations. To reduce the effects of noises, one possible way would be to remove oxides from the contacts as suggested in Ref.~[\onlinecite{Holloway:JAP2013}]. Also, one may think of designing gate operations acting within a nanosecond time scale as investigated theoretically for vortex Majoranas before decoherence takes place \cite{Wu:STA2014, ChingKai:PRR2020}.


\begin{figure*}[!htb]
  \centering
  \includegraphics[scale=0.5]{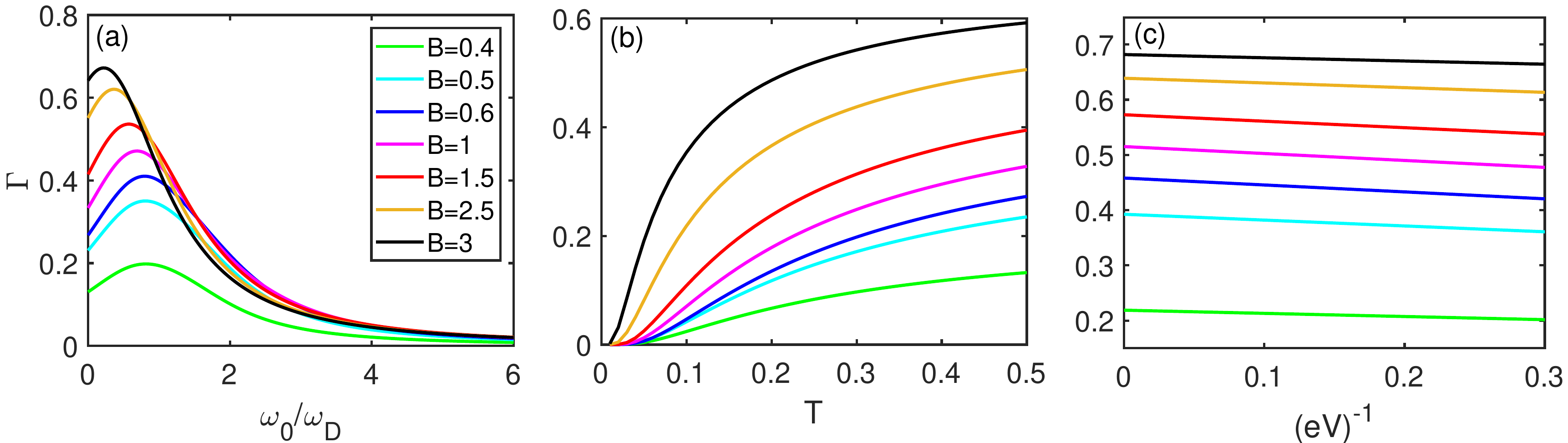}
  \caption{The probability rate of MBSs in a magnetic chain in the presence of (a) Lorentzian, (b) thermal, and (c) QPC noises. In each panel, different curves are for different values of magnetic field $B$. We set $N=48$, $\Delta=0.3$, $J=\mu=5\Delta$ and $\theta=3\pi/5$.}
    \label{fig:KitaevMagnetic}
\end{figure*}

\section{chain of magnetic atoms on a superconductor \label{sec:magnetic}}
The last system we study is a linear chain of magnetic atoms deposited on the surface of an s-wave superconductor, as schematically shown in the right panel of Fig.~\ref{fig:system}. Using the state-of-art spin-polarized scanning tunneling microscopy, it is observed that magnetic chains with more than eight atoms exhibit stable N\'{e}el states which is described by the classical spin model aligned along a local axis \cite{Loth2012}. The magnetic ordering naturally breaks the time-reversal symmetry, and therefore the need for an applied external field is lifted. The magnetic texture induces the effective spin-orbit interaction as electrons move along the chain.

A model Hamiltonian describing the above observation is as follows\cite{NadjPerge2013}:

\begin{eqnarray}
H&=&J\sum_{j,\lambda}a^{\dag}_{j,\lambda}a_{j+1,\lambda} +\sum_{j,\lambda,\lambda^{\prime}}\left[(\vec{B}_{j}\cdot\boldsymbol{\sigma})_{\lambda,\lambda^{\prime}}-\mu\delta_{\lambda,\lambda^{\prime}} \right]a^{\dag}_{j,\lambda}a_{j,\lambda^{\prime}}\cr
&&+\sum_{j}\Delta(a^{\dag}_{j,\uparrow}a^{\dag}_{j,\downarrow}+a_{j,\downarrow}a_{j,\uparrow})
\label{FermionHMagnetic}
\end{eqnarray}
where the magnetic field $\vec{B}_{j}=B\hat{n}_j$ with $\hat{n}_j=(\sin\theta_j\cos\phi_j\hat{x}+
\sin\theta_j\sin\phi_j\hat{y}+\cos\theta_j\hat{z})$ and $\lambda,\lambda^{\prime}$ stand for up and down spin projections. To diagonalize the Hamiltonian we rotate the spins in a local basis with the quantization axis directed along the unit vector $\hat{n}_j$ and without loss of generality assume that $\phi_j=0$, the details of this transformation and subsequent Majorana representation are relegated to Appendix \ref{app:magnetic_chain}.

For Zeeman fields satisfying 
\begin{equation}
\sqrt{\Delta^2+(|\mu|-2|Jf|)^2}<|B|<\sqrt{\Delta^2+(|\mu|+2|Jf|)^2},
\end{equation}
where $f=\cos(\theta/2)$ (see Appendix \ref{app:magnetic_chain}), the superconducting model \eqref{FermionHMagnetic} becomes topologically nontrivial. 

When exposed to dynamical noises, the results of transition rate $\Gamma$ are shown in Fig.~\ref{fig:KitaevMagnetic}. All panels show a qualitatively similar results to nanowire model discussed in preceding section. As seen in Fig.~\ref{fig:KitaevMagnetic}(a) by decreasing the Zeeman field the transition rate is reduced at the lower part of the spectrum. The gap in the quasiparticle spectrum deceases with the rise of the Zeeman field, and the matrix element of MBSs and low lying states increases simultaneously. The cooperation of these two effects gives rise to  the enhancement of the transition rate.

\section{Conclusions}\label{sec:Conclusions}
Before summarizing the main findings of this work, let us recapitulate the main idea and outlines of what we have done. We started by posing an important question of how resilient the MBSs appearing at the open ends of one-dimensional topological superconductors are against dynamical noise sources. We studied the effects of three experimentally relevant time-dependent noises such as Lorentzian, thermal and QPC on MBSs in the Kitaev $p$-wave model, Rashba nanowires, and magnetic atomic chains.    

We showed that in a topological phase the response of MBSs to noise sources depends on the microscopic parameters, and provide a pathway in selecting material combinations where the effects of noises are least. Our findings show that long-range pairings in the Kitaev chain, which can be tunned experimentally, reduce the destructive effects of noises and enhance the robustness of MBSs. For the experimentally realized Rashba nanowire (see Fig.~\ref{fig:estimate} for estimation of transition rates in InAs/Al nanowire heterostructure) and magnetic chain in proximity to an $s$-wave superconductor, which are the most promising proposals for realizing MBSs, we showed that smaller magnetic fields yield more resilience MBSs. In the former case, we showed that the materials with strong Rashba-spin orbit coupling support robust MBSs in a noisy environment.  \\\\

{\it Acknowledgments.}
The authors would like to acknowledge support from the Sharif University of Technology under Grant No. G960208 and the Iran Science Elites Federation.

\appendix

\section{Majorana representation of superconducting nanowire Hamiltonian }\label{app:SOC}
In the basis of the $4N$-components Nambu spinor  $\psi^{\dag}=[\ldots,a^{\dag}_
{j,\uparrow},a^{\dag}_{j,\downarrow},a_{j,\uparrow},a_{j,\downarrow},\ldots]$,
the matrix representation of the Hamiltonian \eqref{SOCMajorana}
can be obtained as \cite{Mazziotti2018}
\begin{equation}
H_{4N\times4N}=
\begin{pmatrix}
H_1 & H_2 & 0 & 0 & 0 &  \\
H_2^{T} & H_1 & H_2 & 0 & \ddots &\vdots \\
0 & \ddots & \ddots & \ddots & \ddots & 0\\
0 & 0 & \ddots & \ddots & \ddots & 0\\
\vdots & \vdots & \ddots & H_2^T & H_1 & H_2\\
0 & 0 & \ldots & 0 & H_2^T & H_1 
\end{pmatrix},
\label{HMatrixSOC}
\end{equation}
where
\begin{equation}
H_1=
\begin{pmatrix}
-\mu & -h & 0 & \Delta \\
-h & -\mu & -\Delta & 0 \\
0 & -\Delta & \mu & h \\
\Delta & 0 & h & \mu
\end{pmatrix}
, H_2=
\begin{pmatrix}
-J & -\alpha & 0 & 0\\
\alpha & -J & 0 & 0 \\
0 & 0 & J & \alpha \\
0 & 0 & -\alpha & J
\end{pmatrix}.
\end{equation}
In the Majorana basis, we use the following unitary transformation:
\begin{equation}
\begin{pmatrix}
c_{2j-1,\uparrow} \\
c_{2j-1,\downarrow} \\
ic_{2j,\uparrow} \\
ic_{2j,\downarrow}
\end{pmatrix}
=U
\begin{pmatrix}
a_{j,\uparrow}\\
a_{j,\downarrow}\\
a^{\dag}_{j,\uparrow}\\
a^{\dag}_{j,\downarrow}
\end{pmatrix}
\text{where}\quad
U=\frac{1}{\sqrt2}\begin{pmatrix}
1 & 0 & 1 & 0\\
0 & 1 & 0 & 1\\
1 & 0 & -1 & 0\\
0 & 1 & 0 & -1
\end{pmatrix}
\label{MajoranaTrans}
\end{equation} 
and rewrite the Hamiltonian in the following form:
\begin{equation}\label{MajSOC}
H=U_{4N\times4N}H_{4N\times4N}
U_{4N\times4N}^{T}.
\end{equation}
that yields $\bf{M}=I_N\otimes(\sigma^x\otimes I _2)$.

\section{Majorana representation of superconducting magnetic chain Hamiltonian}\label{app:magnetic_chain}
Following the strategy used in \cite{NadjPerge2013}, we start with the Hamiltonian \eqref{FermionHMagnetic} and align the spin basis with the unit vector of $\hat{n}_j$ by the following transformation \cite{Choy2011}:

\begin{eqnarray}
\begin{pmatrix}
a_{j,\uparrow}\\
a_{j,\downarrow}
\end{pmatrix}
&=&U_j
\begin{pmatrix}
b_{j,\uparrow}\\
b_{j,\downarrow}
\end{pmatrix}
,\cr\cr\cr
U_j&=&
\begin{pmatrix}
\cos(\theta_j/2) & -\sin(\theta_j/2)e^{-i\phi_j}\\
\sin(\theta_j/2)e^{i\phi_j} & \cos(\theta_j/2)
\end{pmatrix}
\end{eqnarray}

where $b_{j,\lambda}$ satisfies the same anti-commutation relation as $a_{j,\lambda}$.
The Hamiltonian in the new basis reads:
\begin{eqnarray}
H&=&J\sum_{j,\lambda,\lambda^{\prime}}(\Omega_{j,\lambda,\lambda^{\prime}}b^{\dag}_{j,\lambda}b_{j+1,\lambda^{\prime}}+\Omega^{*}_{j,\lambda^{\prime},\lambda}b^{\dag}_{j+1,\lambda}b_{j,\lambda^{\prime}}\cr
&+&B\sigma^z_{\lambda,\lambda^{\prime}}b^{\dag}_{j,\lambda}b_{j,\lambda^{\prime}})-\mu\sum_{j,\lambda}b^{\dag}_{j,\lambda}b_{j,\lambda}\cr
&+&\Delta\sum_{j}(a^{\dag}_{j,\uparrow}a^{\dag}_{j,\downarrow}+a_{j,\downarrow}a_{j,\uparrow}),
\label{rotatedFermionMagnetic}
\end{eqnarray}
where
\begin{equation}
\Omega_j=U^{\dag}_{j}U_{j+1}=
\begin{pmatrix}
f_j & -g^{*}_j\\
g_j & f^{*}_j
\end{pmatrix},
\end{equation}
and
\begin{eqnarray}
f_j=\cos(\theta_j/2)\cos(\theta_{j+1}/2)+\sin(\theta_j/2)
\sin(\theta_{j+1}/2)e^{i(\phi_j-\phi_{j+1})}\cr
g_j=\cos(\theta_j/2)\sin(\theta_{j+1}/2)e^{i\phi_{j+1}}-\sin(\theta_j/2)\cos(\theta_{j+1}/2)e^{i\phi_j}.\cr
\end{eqnarray}
\par
To write the Hamiltonian in the Majorana basis, we use the following definitions
\begin{equation}
b_{j,\lambda}=\frac{1}{2}(c_{2j-1,\lambda}+ic_{2j,\lambda}), \quad 
b_{j,\lambda}^{\dag}=\frac{1}{2}(c_{2j-1,\lambda}-ic_{2j,\lambda}),
\end{equation}
as well as the assumptions of $\phi_j=0$ and the constant angle $\theta$ between nearest-neighbor moments. We define $f_j:=f=\cos(\theta/2)$ and $g_j:=g=\sin(\theta/2)$ and rewrite the Hamiltonian \eqref{rotatedFermionMagnetic} as:
\begin{eqnarray}
H&=&\frac{iJf}{2}\big(c_{2j-1,\uparrow}c_{2j+2,\downarrow}-c_{2j,\uparrow}c_{2j+1,\uparrow}+c_{2j-1,\downarrow}c_{2j+2,\downarrow}\cr
&-&c_{2j,\downarrow}c_{2j+1,\downarrow}\big)-\frac{iJg}{2}\big(c_{2j-1,\uparrow}c_{2j+2,\downarrow}-c_{2j,\uparrow}c_{2j+1,\downarrow}\cr
&-&c_{2j-1,\downarrow}c_{2j+2,\uparrow}+c_{2j,\downarrow}c_{2j+1,\uparrow}\big)-\frac{i\mu}{2}\big(c_{2j-1,\uparrow}c_{2j,\uparrow}\cr
&+&c_{2j-1,\downarrow}c_{2j,\downarrow}\big)+\frac{iB}{2}\big(c_{2j-1,\uparrow}c_{2j,\uparrow}-c_{2j-1,\downarrow}c_{2j,\downarrow}\big)\cr
&+&\frac{i\Delta}{2}\big(c_{2j-1,\downarrow}c_{2j,\uparrow}-c_{2j-1,\uparrow}c_{2j,\downarrow}\big).
\end{eqnarray}
\par
Introducing the following Nambu spinor:
\begin{equation}
\psi^{\dag}=[\ldots,c_
{2j-1,\uparrow},c_{2j-1,\downarrow},ic_{2j,\uparrow},ic_{2j,\downarrow},\ldots],
\end{equation}
the matrix representation of the Hamiltonian reads as \eqref{HMatrixSOC} where
\begin{eqnarray}
H_1&=\frac{1}{2}&
\begin{pmatrix}
0 & 0 & -\mu+B & -\Delta\\
0 & 0 & \Delta & -\mu-B \\
-\mu+B & \Delta & 0 & 0 \\
-\Delta & -\mu-B & 0 & 0 
\end{pmatrix},
\cr
H_2&=\frac{1}{2}&
\begin{pmatrix}
0 & 0 & f t & -g t\\
0 & 0 & g t & f t\\
f t & -g t & 0 & 0\\
g t & f t & 0 & 0
\end{pmatrix},
\end{eqnarray}
leading to $\bf{M}=I_N\otimes(\sigma^x\otimes I _2)$.

\vspace*{-0.25in}

%

\end{document}